\documentclass[
reprint,
amsmath, 
amssymb, 
aps, 
superscriptaddress, 
pra]{revtex4-2}

\usepackage{graphicx}
\usepackage{dcolumn}
\usepackage{bm}
\usepackage{braket} 
\usepackage{hyperref}
\usepackage{subfigure} 
\usepackage{nccmath}
\usepackage{amsthm} 
\usepackage{mathrsfs} 
\usepackage{xfrac} 
\usepackage{pifont}
\usepackage{adjustbox}
\usepackage{bbm} 
\usepackage[utf8]{inputenc} 
\usepackage[normalem]{ulem}
\usepackage[usenames,dvipsnames,svgnames,table]{xcolor}
\usepackage[sort&compress]{natbib}
\usepackage{color}
\usepackage{longtable}
 
\usepackage{commath}

\begin{document}
\title{Decoherence predictions in a superconductive quantum device using the steepest-entropy-ascent quantum thermodynamics framework}

\author{J. A. Monta\~nez-Barrera}
 	\altaffiliation{ja.montanezbarrera@ugto.mx (J.A. Monta\~nez-Barrera)}
	\affiliation{Department of Mechanical Engineering, Universidad de Guanajuato, Salamanca, GTO 36885, Mexico}

\author{Michael R. von Spakovsky}
	\altaffiliation{vonspako@vt.edu (M.R. von Spakovsky)}
	\affiliation{Department of Mechanical Engineering, Virginia Tech, Blacksburg, VA 24061, USA}		

\author{Cesar E. Damian Ascencio}
 	\altaffiliation{cesar.damian@ugto.mx (C.E. Damian-Ascencio)}
	\affiliation{Department of Mechanical Engineering, Universidad de Guanajuato, Salamanca, GTO 36885, Mexico}

\author{Sergio Cano-Andrade}
	\altaffiliation{sergio.cano@ugto.mx (S. Cano-Andrade)}
	\affiliation{Department of Mechanical Engineering, Universidad de Guanajuato, Salamanca, GTO 36885, Mexico}

\begin{abstract}
The current stage of quantum computing technology, called noisy intermediate-scale quantum (NISQ) technology, is characterized by large errors that prohibit it from being used for real applications. In these devices, decoherence, one of the main sources of error, is generally modeled by Markovian master equations such as the Lindblad master equation. In this work, the decoherence phenomena are addressed from the perspective of the steepest-entropy-ascent quantum thermodynamics (SEAQT) framework in which the noise is in part seen as internal to the system.  The framework is as well used to describe changes in the energy associated with environmental interactions. Three scenarios, an inversion recovery experiment, a Ramsey experiment, and a two-qubit entanglement-disentanglement experiment, are used to demonstrate the applicability of this framework, which provides good results relative to the experiments and the Lindblad equation, It does so, however, from a different perspective as to the cause of the decoherence. These experiments are conducted on the IBM superconductive quantum device $ibmq\_bogota$.

\begin{description}
	\vspace{0.2cm}
	\item[Keywords] IBMQ; Qiskit; Quantum Computation; Entanglement and Correlations; Steepest-Entropy-Ascent.
\end{description}

\end{abstract}

\maketitle

\section{\label{int}Introduction}

\hspace{1em}Decoherence, which is perhaps one of the most critical aspects of quantum computation, is the loss of information that exists in the subsystems of a quantum device. It is typically viewed as resulting from environmental effects and random disturbances that affect the capacity of quantum systems to store information. Hence, the development of realistic quantum computers requires understanding, controlling, and/or correcting for decoherence.

\hspace{1em}The typical approach to modeling decoherence is to use linear Markovian quantum master equations (QMEs) of the Kossakowski-Lindblad-Gorini-Sudrashan type to represent the dynamics of system state evolution \cite{Lindblad1976, Nakatani2010, Chou2008} and the loss of correlation. These QMEs assume that the system interacts with an environment and that the only relevant effect is that on the system. Nevertheless, the QMEs are still linear in nature and, thus, can at best only mimic the non-linear dynamics that may be in play. Despite this fact, QMEs have shown good agreement with experimental data \cite{Turchette2000, Stickler2018, Barnes2012}. Even so, if the weak interactions needed for the QMEs equations are real, Nakatani and Ogawa \cite{Nakatani2010} have shown that the Born-Markov approximation for obtaining evolution equations, i.e., quantum master equations (QMEs), cannot be used for composite systems in the strong-coupling regime, no matter how short the reservoir correlation time.

\hspace{1em}An alternative approach results when instead of assuming that the relevant irreversible effect on the system is due to an environment, quantum mechanics (QM) is complemented with the second law of thermodynamics represented by the steepest entropy ascent (SEA) principle. Such an approach assumes that the irreversible effect is fundamental to the system itself. This idea can be traced back to the work of Hatsopoulos and co-workers \cite{Hatsopoulos1976, Hatsopoulos1976-1, Hatsopoulos1976-2, Hatsopoulos1976-3, Beretta1984} and has matured over the last four decades and  grown substantially in the last decade. A consequence of this work has been the construction of dynamical models based on the SEA principle that explain non-equilibrium phenomena at all levels of description, from the macroscopic to the microscopic (e.g., \cite{Beretta2006, Beretta2009, Smith2012,  VonSpakovsky2014, Montefusco2015, Cano2015,Montanez-Barrera2020, Li2018, Li2017, Li2016, Li2015, Li2014, Kusaba2017, Kusaba2019, Yamada2018, Yamada2019, Yamada2019a, Yamada2019b, Yamada2020, Goswami2021, McDonald2021, Montanez-Barrera2021}).

\hspace{1em}With the advent of NISQ devices \cite{Preskill2018}, the study and simulation of noise in quantum devices have attracted great attention. Applications range from error mitigation techniques \cite{Kandala2019, Sagastizabal2019, Smith2021} to the simulation of quantum devices with real noise \cite{Murali2019, Noh2020, Dahlhauser2021} to the simulation of quantum algorithms with decoherent error to understand how this noise affects NISQ algorithms in quantum machine learning (QML) models \cite{Wang2021}. Recent work suggests that the SEAQT  framework is suitable for modeling decoherence in quantum computation as is shown in \cite{Cano2015} where SEAQT is used to show the interaction of a quantum cavity with a qubit and in \cite{Montanez-Barrera2020} where it is used to predict the state evolution of a two-qubit system when a control-phase gate is applied to a double quantum dot architecture. Both models are compared with experiments and show good predictive capabilities.

\hspace{1em}In the present work, the loss of coherence in three different scenarios is experimentally evaluated. First, a qubit relaxation experiment is implemented to determine how fast the system looses information due to the interaction with the environment. This phenomenon is measured in terms of the time $T_1$. Next, the dephasing on individual qubits using the Ramsey experiment is evaluated. This experiment measures the loss of phase by a system through time, which is quantified with the time $T_2^*$. The last experiment is a two-qubit cross-resonance interaction. Here, the loss of entanglement by a composite system is evaluated. To conduct this experiment, a two-qubit system in a Bell state $|\Phi\rangle$ is entangled and disentangled. The experiments are conducted on IBM's $ibmq\_bogota$ quantum device in qubits 0 to 4 for the $T_1$ and $T_2^*$ experiment and in qubits 0 and 1 for the entanglement experiment. The experimental results are compared with simulations using the SEAQT framework and the Lindblad approach.

\hspace{1em}The paper is organized as follows. In Sections \ref{SecIIA} to \ref{SecIID} different features of the SEAQT equation of motion are laid out and discussed, while Section \ref{SecIIE} presents the Lindblad type quantum master equation used here. Section \ref{SecIII} then describes the experimental setup of IBM's $ibmq\_bogota$ quantum device for each of the three experiments conducted. Section \ref{SecIV} then provides the results of the experiments and the simulations and a discussion of the results. The paper then wraps up with a number of conclusions in Section \ref{SecV}.

\section{Mathematical Model}\label{SecII}
\subsection{The SEAQT framework}\label{SecIIA}
\hspace{1em} In the SEAQT framework, the dynamics of the density operator, $\hat \rho$, of a quantum system is governed by both a symplectic (unitary) and a dissipation (non-unitary) term. The former, the so-called von Neumann term of quantum mechanics, captures the reversible (i.e., linear) dynamics of state evolution, while the latter, which is based on the principle of steepest entropy ascent (SEA), captures the irreversible (i.e., nonlinear) dynamics. This principle states that at every instant of time the density operator evolves in the direction of maximal entropy increase such that the conservation constraints placed on the generators of the motion (e.g., the Hamiltonian and the identity operator) are satisfied.  Note that the view of physical reality assumed here is one in which the nonlinear dynamics of state evolution resulting from the dephasing phenomenon are intrinsic to the system and not a consequence of interactions with an environment. This contrasts with the standard open quantum system
framework (see Section \ref{SecIIE}) that forms the basis for the Lindblad equation of motion, which assumes that this phenomenon is the result of a continual cyclic buildup and loss of correlations between the system and environment. As pointed out in Section \ref{int}, this assumption fails if the system-environment coupling is strong \cite{Nakatani2010}, a limitation which does not apply to the SEAQT framework. Of course, in the case of the relaxation phenomenon, the SEAQT framework also assumes a system-environment interaction since this phenomenon involves an exchange of energy between the system and environment. However, there is no limitation in the SEAQT framework on the strength of the coupling.

\hspace{1em} The SEAQT equation of motion for a general quantum system \cite{Beretta1985} is written as

\begin{equation}\label{Eq:Beretta}
\frac{d\hat{\rho}}{dt} = -\frac{i}{\hbar}[\hat{H},\hat{\rho}] - \sum_{J} \left(  \frac{1}{\tau_{D_J}} \hat{D}_J\otimes \hat{\rho}_{\bar J}\right)
\end{equation}
where $\hat H$ and $\hat\rho$ are the Hamiltonian and the density operator, respectively, for a composite system. The $\hat\rho_J (J = 1, 2,...)$ are the density operators for each individual qubit with $\hat \rho_J = \mathrm{Tr}_{\bar J}(\hat \rho)$ and $\bar J$ indicating the direct product on the Hilbert space that does not contain the subsystem $J$. In addition, the $\tau_{D_J} (J = 1, 2,...)$ are internal-relaxation parameters that are positive constants or positive functionals of the $\hat\rho_J$, while the $\hat D_J (J = 1, 2...)$ are the dissipation operators for each qubit. To assure positivity and hermiticity of the density operator, $\hat \rho$, the latter operators are written as

\begin{equation}\label{D_J}
\hat{D}_J = \frac{1}{2}\left( \sqrt{\hat{\rho}_J} \tilde{D}_J + (\sqrt{\hat{\rho}_J}\tilde{D}_J)^{\dagger}\right) 
\end{equation}
where the symbol $\dag$ signifies the adjoint and each $\tilde D_J$ for a two-qubit system is expressed as

\begin{equation} \label{EqtildeD_J}
\tilde{D}_J = \frac{
\begin{vmatrix}
\sqrt{\hat{\rho}_J}(\hat{B}\ln{\hat{\rho}})^J & \sqrt{\hat{\rho}_J}(\hat{I})^J & \sqrt{\hat{\rho}_J}(\hat{H})^J \\
(\hat{I},\hat{B} \ln{\hat{\rho}})^J &(\hat{I},\hat{I})^J&(\hat{I},\hat{H)}^J\\
(\hat{H},\hat{B} \ln{\hat{\rho}})^J&(\hat{H},\hat{I})^J&(\hat{H},\hat{H})^J\\
\end{vmatrix}
}{
\begin{vmatrix}
(\hat{I},\hat{I})^J & (\hat{I},\hat{H})^J\\
(\hat{H},\hat{I})^J & (\hat{H},\hat{H})^J\\
\end{vmatrix}
} .
\end{equation}
Here $(\cdot,\cdot)^J$ is the Hilbert-Schmidt inner product defined on Hilbert space $\mathcal{H}^J$ by $(\hat{F},\hat{G})^J=\textrm{Tr}_J(\hat{\rho}_J\{(\hat{F})^J,(\hat{G})^J\})$ with $J=A$, $B$, $(\hat{F})^A=\textrm{Tr}_B[(\hat{I}_A\otimes\hat{\rho}_B)\hat{F}]$, and $(\hat{F})^B=\textrm{Tr}_A[(\hat{\rho}_A\otimes \hat{I}_B)\hat{F}]$. In Eq. (\ref{EqtildeD_J}), $\hat{B}$ is the projector onto the range of $\rho$, i.e., the idempotent operator that results from summing up all of the eigenprojectors of $\hat{\rho}$ belonging to its nonzero eigenvalues. For more details, the reader is referred to \cite{beretta1984quantum,beretta2010maximum}. 

\subsection{Equation for a system interacting with a reservoir}\label{SecIIB}
\hspace{1em}One way to represent the interaction between a system $J$ and a reservoir $R$ consists of considering the degrees of freedom of both subsystems to be orthogonal \cite{Holladay2019}. This allows one to represent the Hilbert space of the two subsystems  as $\mathcal{H} = \mathcal{H_S}\oplus\mathcal{H_R}$. In this framework, the eigenenergies of both subsystems are independent and unentangled. The representation of the dissipative term for the system is 

\begin{equation} \label{EqtildeD_S}
\tilde{D}_{JR} = -\sqrt{\hat{\rho}_J}\frac{
\begin{vmatrix}
-\hat{B}\ln{\hat{\rho}_J} & \hat{I}_J & \hat 0_J & \hat H_J\\
\left< s\right>_J & P_J & 0 & \left< e\right>_J\\
\left< s\right>_R & 0 & P_R & \left< e\right>_R\\
\left< es\right>_J + \left< es\right>_J& \left< e\right>_J & \left< e\right>_R & \left< e^2\right>_J + \left< e^2\right>_R\\
\end{vmatrix}
}{
\Gamma
} .
\end{equation}
Here the expected values for the system are $\langle s \rangle_{J} = -\textrm{Tr} (\hat \rho_{J} \ln \hat \rho_{J})$, $\langle e \rangle_{J} = \textrm{Tr} (\hat \rho_{J} \hat H_{J})$, and $\langle es \rangle_{J} = - \textrm{Tr} (\hat \rho_{J} \hat H_{J} \ln \hat \rho_{J})$. Also, $|\cdot|$ is a determinant and $\Gamma$ is a Gram determinant. Expanding this last expression results in

\begin{equation}
\tilde{D}_{JR} = \sqrt{\hat{\rho}_J} \left( \hat{B}\ln{\hat{\rho}_J} - \frac{B_1}{\Gamma} \hat I_J - \frac{B_3}{\Gamma}\hat H_J \right)
\end{equation}
where

\begin{equation}
B_1 = 
\begin{vmatrix}
\left< s\right>_J & 0 & \left< e\right>_J\\
\left< s\right>_R & P_R & \left< e\right>_R\\
\left< es\right>_J + \left< es\right>_R & \left< e\right>_R & \left< e^2\right>_J + \left< e^2\right>_R\\
\end{vmatrix},
\end{equation}

\begin{equation} 
B_3 =
\begin{vmatrix}
\left< s\right>_J & P_J & 0 \\
\left< s\right>_R & 0 & P_R \\
\left< es\right>_J + \left< es\right>_R& \left< e\right>_J & \left< e\right>_R\\
\end{vmatrix},
\end{equation}

\begin{equation} \label{EqtildeD_S}
\Gamma =
\begin{vmatrix}
P_J & 0 & \left< e\right>_J\\
0 & P_R & \left< e\right>_R\\
\left< e\right>_J & \left< e\right>_R & \left< e^2\right>_J + \left< e^2\right>_R\\
\end{vmatrix}.
\end{equation}
In the limit, when the number of eigenlevels of the reservoir $P_R$ are much greater than those of the system $P_J$ ($P_R \ll P_J$), one can show that $B_3$ reduces to

\begin{equation}
\frac{B_3}{\Gamma} \approx \frac{\left< es \right>_R -  \left< e\right>_R \left< s\right>_R}{ \left< e^2\right>_R -  \left< e\right>_R^2}.
\end{equation}

\hspace{1em}Now assuming a canonical distribution for the reservoir characterized by the inverse temperature $\beta_R$ and the Hamiltonian $\hat H_R$, $B_3$ is approximately

\begin{equation}
\frac{B_3}{\Gamma} \approx -\beta_R.
\end{equation}
Therefore, $\tilde{D}_J$ of Eq. (\ref{EqtildeD_J}) for the system (i.e., for $J$) interacting with a reservoir can be expressed as

\begin{equation} \label{EqtildeD_S}
\tilde{D}_{JR} = \sqrt{\hat{\rho}_J} \left( \hat{B}\ln{\hat{\rho}_J} + \left< s \right>_J \hat I_J + \beta_R(\hat H_J - \left< e \right>_J \hat I_J) \right).
\end{equation}
The first two terms in Eq. (\ref{EqtildeD_S}) account for dephasing in the system, while the last term, i.e., $\beta_R(\hat H_J - \left< e \right>_J \hat I_J)$, accounts for relaxation.

\subsection{Model of relaxation and dephasing}\label{SecIIC}

\hspace{1em}
Even though Eq. (\ref{EqtildeD_S}) by its own can represent the phenomena of relaxation and dephasing, experimental results show that those phenomena are happening at different rates. This can be taken into account by increasing or decreasing the value of $\beta_R$ in Eq. (\ref{EqtildeD_S}) relative to the rate at which the relaxation occurs or, alternatively, by separating the effects of the phenomenon of relaxation from that of dephasing as in the Lindblad equation. Therefore, in the SEAQT framework, it is coupled the phenomenon of dephasing with that of relaxation, the general form of the SEAQT equation of motion, Eq. (\ref{Eq:Beretta}), is modified to include the effects of the reservoir interaction that only generates the relaxation transition i.e., with $\beta_R(\hat H_S - \left< e \right>_S \hat I_S)$ in Eq. (\ref{EqtildeD_S}). Therefore, the equation of motion for this case is written as 

\begin{equation}\label{Eq:BerettaRes}
\frac{d\hat{\rho}}{dt} = -\frac{i}{\hbar}[\hat{H},\hat{\rho}] - \sum_{J} \left(  \frac{1}{\tau_{D_J}} \hat{D}_J\otimes \hat{\rho}_{\bar J} + \frac{1}{\tau_{D_R}} \hat{D}_{JR} \otimes \hat{\rho}_{\bar J}\right)
\end{equation}
where 
\begin{equation}
\hat{D}_{JR} = \frac{1}{2}\left( \sqrt{\hat{\rho}_J} \tilde{D}_{JR} + (\sqrt{\hat{\rho}_J}\tilde{D}_{JR})^{\dagger}\right)
\end{equation}
and
\begin{equation}
\tilde D_{JR} = \sqrt{\hat{\rho}_J} \left( \beta_R(\hat H_J - \left< e \right>_J \hat I_J) \right).
\end{equation}

\subsection{Qubit-reservoir interaction relaxation parameter}\label{SecIID}

\hspace{1em}A usual approach for simulating a quantum process with the SEAQT equation of motion is to consider the relaxation parameter $\tau_D$ in Eq. (\ref{Eq:Beretta}) as a constant determined for a specific process. This approach has given good results when the process is at constant energy. However, the energy in the one-qubit experiments described in this paper changes considerably with time. A link between the rate of change of quantum states and the energy of the system proposed by Mandelstam and Tamm \cite{Mandelstam1945, DelCampo2013} shows that the quantum speed limit can be bound by the energy of the system. In addition, Fermi's Golden rule \cite{Braak2020}, which describes the transition rate between quantum states of a quantum system as a result of a weak perturbation, also describes such a transition in terms of the energy of the system. Thus, it is assumed here that the relaxation parameter fluctuates with the energy of the system. In particular, a qubit-reservoir relaxation parameter $\tau_{D_R}$ varying linearly with the expectation energy is assumed such that

\begin{equation}
\tau_{D_R}(\hat{\rho}(t)) = x_0 (1 + \langle\hat H\rangle) 
\end{equation}
where $x_0$ is a constant to be determined. The Hamiltonian for a transmon qubit can be described using a Duffing oscillator \cite{Khani2009} so that

\begin{equation}
\hat H = \omega \hat{b}^\dagger \hat{b} + \frac{\delta}{2} \hat{b}^\dagger \hat{b} (\hat{b}^\dagger \hat{b} - \hat{I})
\end{equation}
where $\omega$ and $\delta$ are the transmon frequency and anharmonicity, respectively, and $\hat{b}$ is the annihilation operator. Using the definitions that $\hat{b}^\dagger \hat{b} = \sum_j j \left|j\right>\left<j\right|$ for the eigenlevels $j$ of the transmon and $\omega_j = \left(\omega - \frac{\delta}{2}\right) j + \frac{\delta}{2}j^2 $, the Hamiltonian is rewritten as

\begin{equation}
\hat H = \sum_j\omega_j \left|j\right>\left<j\right|.
\end{equation}

For simplicity, it is assumed that the transmon is a two-level system. In that case, the Hamiltonian is given by

\begin{equation}
\hat H = -\frac{1}{2}\hbar \omega_q\hat{\sigma}_z
\end{equation}
and as a consequence, the relaxation parameter can be written as 

\begin{equation}\label{tauDR}
\tau_{D_R}(\hat{\rho}) = x_0 (1 + \textrm{Tr}(\hat\rho(t) \hat\sigma_z)).
\end{equation}
Here, $\hat{\sigma}_z$ is the z-component Pauli matrix with the set of Pauli matrices given by $\hat\sigma_{x} = \begin{bmatrix}0&1\\1&0\end{bmatrix}$, $\hat\sigma_{y} = \begin{bmatrix}0&-i\\i&0\end{bmatrix}$, and $\hat\sigma_{z} = \begin{bmatrix}1&0\\0&-1\end{bmatrix}$.

\subsection{The open quantum system model}\label{SecIIE}

\hspace{1em}The Lindblad equation, which is also known as the Gorini–Kossakowski–Sudarshan–Lindblad (GKLS) master equation, predicts the evolution of state of a quantum system as Markovian interactions between the system and multiple baths \cite{Lindblad1976}. Generally, it uses a linear description to predict the non-linear evolution of the density operator $\hat \rho$, preserving the laws of quantum mechanics and assuming a weak interaction between the system and the environment (baths). It has played an important role in quantum information and decoherence \cite{Lidar1998, Kraus2008, Brun2000, Schlosshauer2019}, which makes it suitable for the present study. The equation of motion of the Lindblad type used here is expressed as

\begin{equation}\label{eq8}
\frac{d \hat{\rho}_s}{dt} = -\frac{i}{\hbar} [\hat{H},\hat{\rho}_s] + \frac{1}{2} \displaystyle\sum_{j=1}^2\gamma_j(  2\hat{L}_j \hat{\rho}_s \hat{L}_j^\dagger - \hat{L}_j^\dagger \hat{L}_j \hat{\rho}_s  -  \hat{\rho}_s  \hat{L}_j^\dagger  \hat{L}_j)
\end{equation}
where

\begin{equation}
\hat{L}_1 = \sqrt{\gamma_1} \hat{b}
\end{equation}
and 

\begin{equation}
L_2 = \sqrt{\gamma_2} \hat{\sigma}_z.
\end{equation}
The $L_1$ operator is used to simulate the amplitude damping phenomenon (i.e., the relaxation), while the $L_2$ operator is employed to simulate dephasing. $\gamma_{1}$ is the strength of the relaxation and $\gamma_2$ is that of the dephasing.

\section{Experimental setup}\label{SecIII}

\hspace{1em}The $ibmq\_bogota$ device used in the experiments is a superconductive quantum processor with a Falcon r4L architecture, 5 qubits, a quantum volume (QV) of 32, an average $T_1=$ 88.34 $\mu s$, an average $T_2=$ 128.54 $\mu s$, and an average CNOT gate (see the Appendix) error of 1.056 x 10$^{-2}$. Qubits q0 to q5 are employed with the following excitation frequencies: $f_{0,1,2,3,4} = 5.00, 4.85, 4.78, 4.86, 4.98$ GHz. Three experiments were conducted on this device: an inversion recovery experiment, a Ramsey experiment, and a two-qubit entanglement experiment in the qubits 0 and 1. The first two experiments involved all of the qubits{\color{red}.} The inversion recovery and the Ramsey experiments are based on the Hamiltonian given by
		
\begin{equation}
	\hat H = \hbar \Delta\omega\hat\sigma_z + \hbar \theta_G(t)\hat\sigma_y
	\label{ham}
\end{equation}
		where $\Delta\omega = \omega_q - \omega_d$, $\omega_q$ and $\omega_d$ are the qubit and electric field frequencies, respectively, and $\theta_G(t)$ is the pulse applied to the qubit to generate a transition, a $\pi$ rotation for the inversion recovery experiment, and a $\pi/2$ rotation for the Ramsey experiment.

\hspace{1em}For all three experiments, the number of shots, $n$, for each experiment is 8192 and is the number of times that an experiment is repeated to get back the probability of $\langle \hat{Z} \rangle=\textrm{Tr}\left(\hat{\rho}\hat{\sigma}_z\right)$ on each qubit. The inversion recovery and the Ramsey experiments are repeated 4 times with some time lapse between each experiment. Their variation is presented in the results with the points representing the mean value and the error bars the standard deviation. This number of shots is the maximum number allowed and reduces the statistical error of the experiment to $1/\sqrt{n}$.

\subsection{Inversion Recovery}
\begin{figure}[htbp!]
\includegraphics[width=6cm]{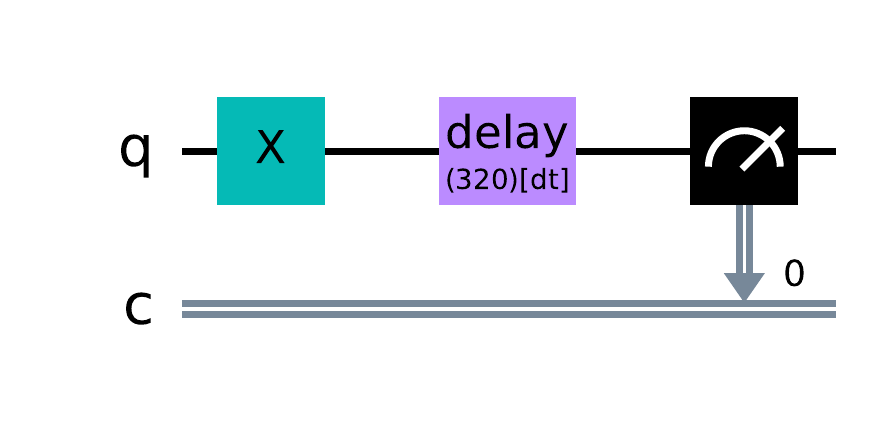}\\
\centering a) \\
\includegraphics[width=9cm]{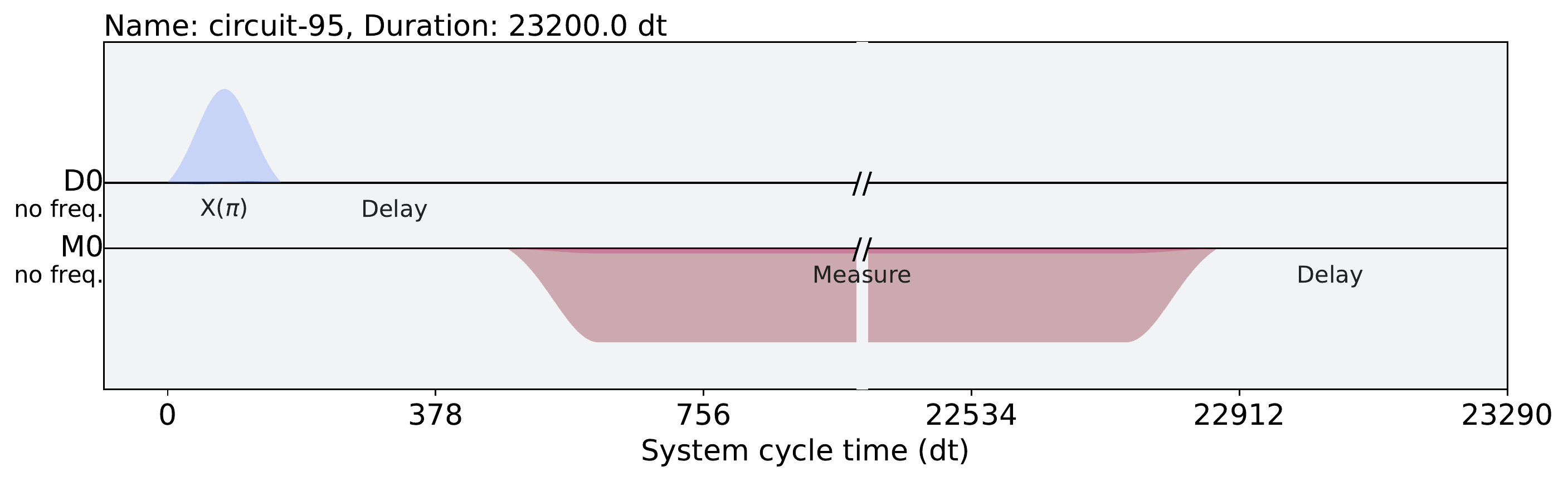}\\
\centering b) \\ 

\caption{\label{t1} Schematic representation of an inversion recovery experiment characterized by $T_1$: a) the pulse in the D0 channel is a $X_\pi$ gate, M0 is the measurement, and b) a pulse is applied in this channel to recover the state after the delay time. The experiments were conducted on the $ibmq\_bogota$ device from IBM, and using an open-pulse control in the Qiskit python library.}
\end{figure}

\hspace{1em}The inversion recovery experiment provides information about how fast a qubit suffers thermalization because of its interaction with an environment (or reservoir). It is measured with the relaxation time $T_1$ where $T_1$ is the time that it takes the qubit to reach $1 - 1/e$ or 63\% of its initial condition. Here, a $\pi$ pulse gate (X gate) is used to move the system from state $\left|0\right>$ to state $\left|1\right>$. Next, a delay time is applied and measurement made. This process is repeated with 25 different delay times, ranging from 0 to 42.6 $\mu s$. What is observed is a decay of the probability of being in state $|1\rangle$. This decay can be approximated with the relation,

\begin{equation}
\hat{\rho}(t) = \hat{\rho}(0) (1 - e^{-t/T_1})
\end{equation}   
where $\hat{\rho}(0)$ is the density operator at time 0, $t$ is the time, and $T_1$ is the relaxation time constant. Fig. \ref{t1} schematically shows the circuit and the pulse representation of this experiment. Here, D0 is the channel, which transmits the signals to the qubits, allowing single qubit gate operations; and M0 is a measurement channel, which transmits a measurement stimulus pulse for readout. The pulse on D0 is an X gate. This $X_{\pi}$ gate's pulse is known as a derivative removal via an adiabatic gate (DRAG) \cite{Motzoi2009} composed of a Gaussian shaped pulse $\theta_G(t) = \frac{1}{\sqrt{2\pi}\sigma}e^{-(\frac{t - t_o}{\sigma})^2}$ with rotation $\pi$ about the y-axis and a derivative of the Gaussian pulse responsible for eliminating $X_{\pi}$ imperfections about the x-axis. The $X_{\pi}$ has a gate time of $t_g = 35.2 \ ns$ where $\sigma = t_g/4$.

\subsection{Ramsey experiment}
\begin{figure}[htbp!]
\includegraphics[width=6cm]{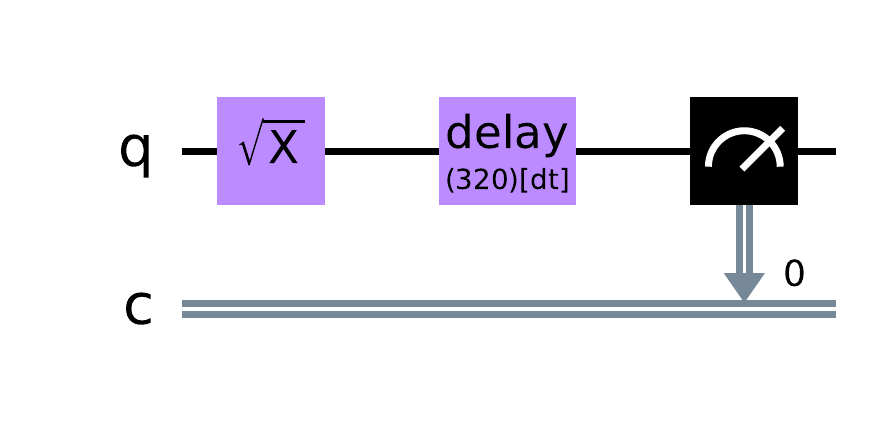}\\
\centering a)\\
\includegraphics[width=9cm]{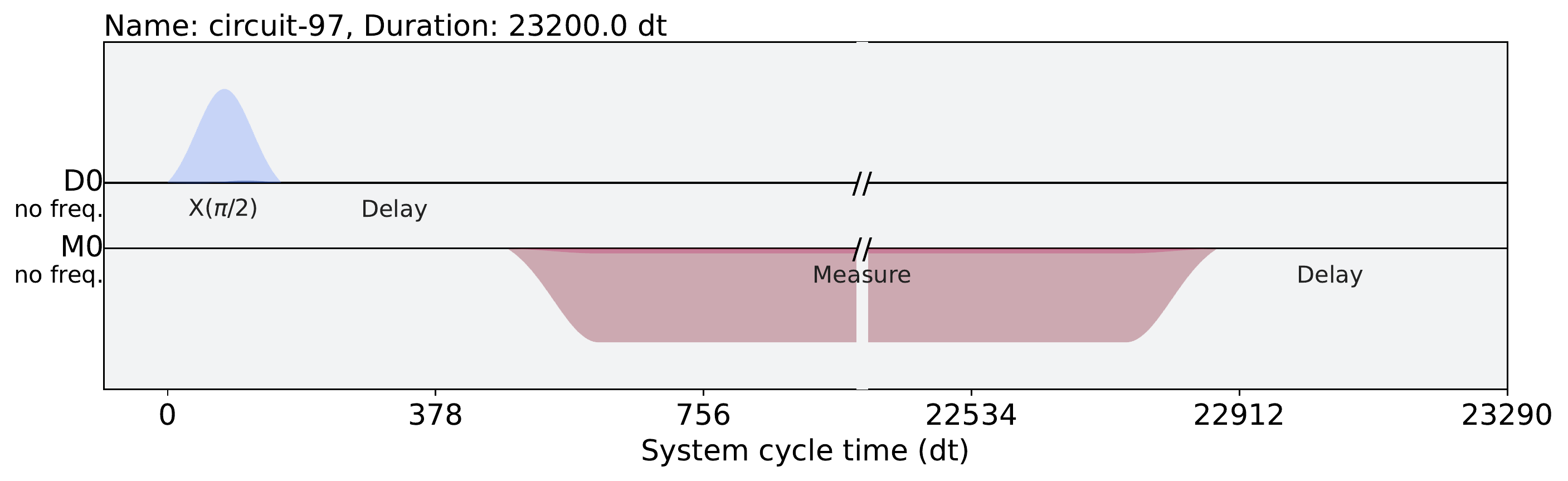}\\
\centering b)
\caption{\label{t2} Schematic representation of the Ramsey experiment a) circuit mode and b) pulse mode to determine the coherence time $T_2^*$: the experiment is conducted using the $ibmq\_bogota$ device and the pulse level control from IBMQ. }
\end{figure}

\hspace{1em}The Ramsey experiment measures the dephasing time $T_2^*$ and the qubit detuning \cite{Paik2011}. Ideally, the frequency used for the pulse rotations is the resonant frequency of the qubit. However, due to imperfections and an inability to tune the resonant frequency, the qubit suffers from an oscillation proportional to the detuning. On the other hand, the dephasing phenomenon moves the qubit's Bloch vector towards the center of the Bloch sphere. The experiment consists in applying a $X_{\pi/2}$ gate pulse on the drive channel D0 and then allowing the system to evolve during a delay time, after which the qubit $\langle \hat Z \rangle$ observable is measured. Fig. \ref{t2} presents the circuit and pulse representation for the experiment to determine $T_2^*$ relative to qubit 0.

\subsection{Two-qubit entanglement state experiment}

\begin{figure}[htbp!]
\hspace*{-0.5cm}\includegraphics[width=8.7cm]{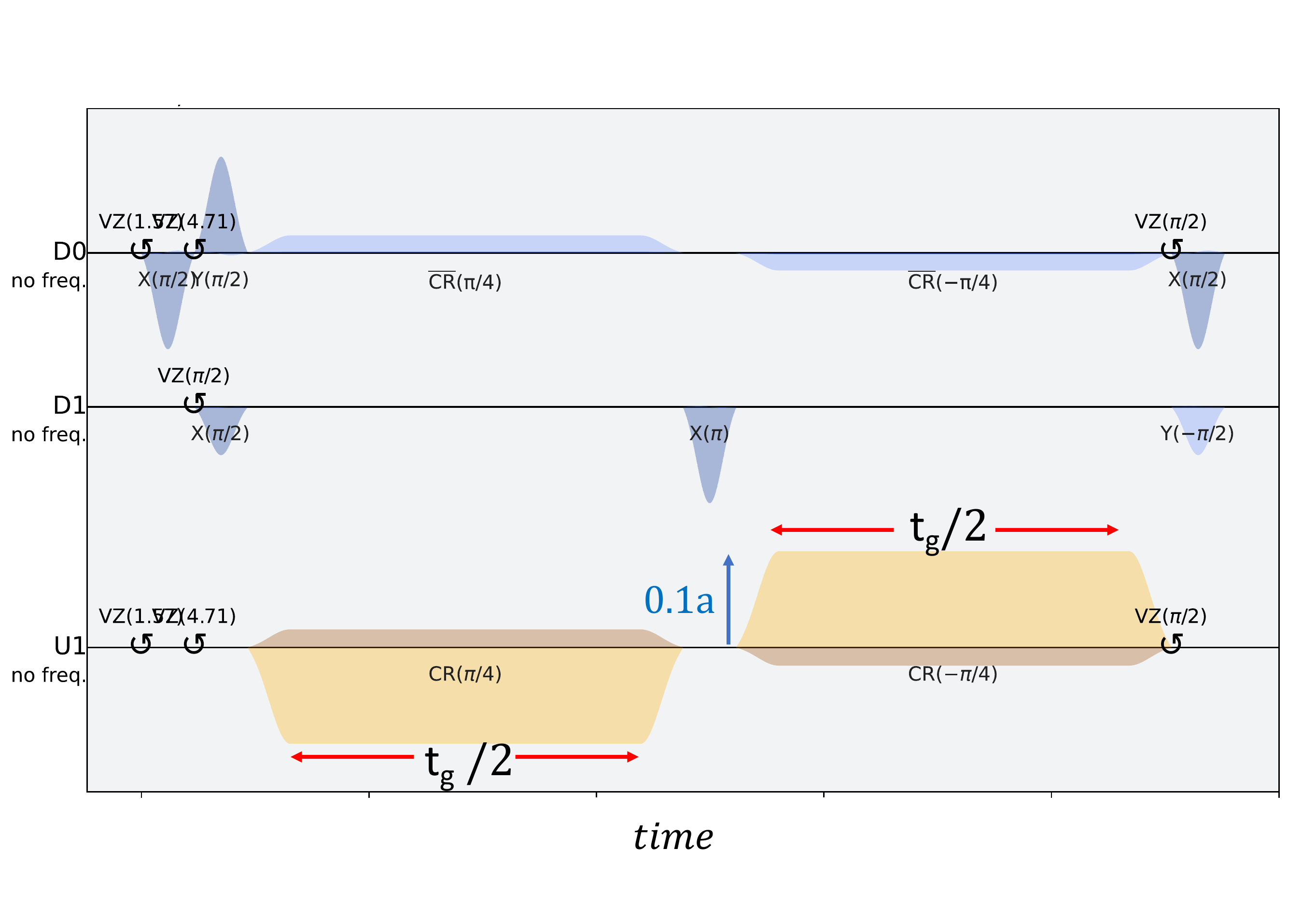}
\caption{\label{Bell}  The circuit to create the entanglement and disentanglement sequence: here, $t_g/2$ is the width of the CR pulse and 0.1a corresponds to the amplitude of the pulse where `a' is the default amplitude of the CR pulse for  a CNOT gate. The channel D0 represents the pulses on the control qubit and D1 the pulses on the target qubit and the yellow pulses in U1 represent the CR pulses that entangle both qubits.}
\end{figure}

\hspace{1em}The two-qubit entanglement state experiment, shown in Fig. \ref{Bell}, consists of an entanglement and disentanglement scenario where the Bell state $\left|\Phi\right> = 1/\sqrt{2}(|00\rangle + |11\rangle)$ is obtained in the case of maximum entanglement. This is achieved using a Hadamard gate on the control qubit followed by a cross resonance (CR) protocol with a Han echo sequence on the CR channel \cite{Hahn1950,Sheldon2016}. This sequence is used to reduce the noise of the CR channel due to imperfections of the pulse applied. This is explained in detail in the Appendix. In Fig. \ref{Bell}, D0 and D1 are the drive channels for Q0 and Q1, respectively, and U1 is the control channel for the interaction between Q0 and Q1. A modification of the default $ibmq\_bogota$ pulse calibration is used with a change in the U1 and D0 amplitude to 0.1 of the default pulses during the CR section. In addition, the CR pulse width is modified from $t_g = 0$ to $t_g = 20.45$ $\mu$s with 30 intermediate pulse widths. Here, a maximum time of 20.45 $\mu s$, which is different from the time of the 1 qubit experiments, is used because there is a limitation on the number of samples that can be created for a pulse in the IBM quantum hardware. In this case, that limit  is closed to 20.45 $\mu s$ for the CR pulse. Furthermore, for the one-qubit experiments, a delay time, which does not involve a sample pulse, is used. To construct the density state operator,  a tomography process based on the work of Smolin {\it et al.} \cite{Smolin2018} is used with 9 independent measurements to recover the two-qubit system density operator.

\section{Results}\label{SecIV}
\subsection{Inversion Recovery Experiment}
\begin{figure}[htbp!]
\includegraphics[width=9cm]{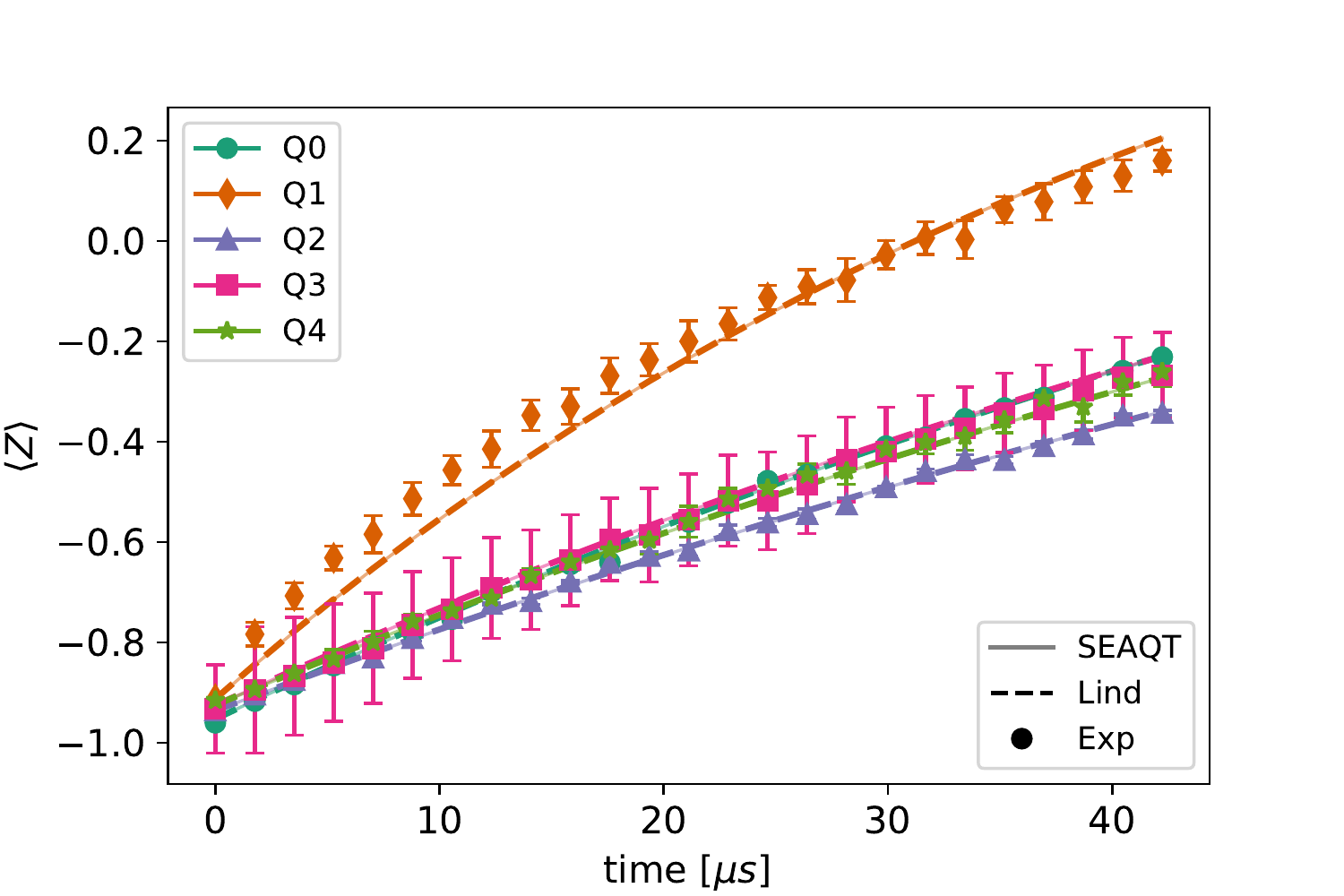}
\caption{\label{t1_res} Results from the inversion recovery experiment for the time evolution of the $\langle \hat{Z} \rangle$ component or observable: the experimental results for all five qubits are compared with simulation results from the SEAQT and the Lindblad equations of motion. The error bars represent the standard deviation.}
\end{figure}

\hspace{1em}The inversion recovery experiment is used to characterize how fast a qubit loses information because of an interaction with the environment. Fig. \ref{t1_res} shows the experimental results of the $\langle \hat{Z} \rangle$ component for the 5 qubits of $ibmq\_bogota$. As seen, the probability of getting state $|1 \rangle$ monotonically decreases with time until it reaches a point close to the $|0\rangle$ state. This phenomenon is modeled using the Lindblad equation employing an annihilation operator in the master equation. To model this phenomenon within the SEAQT framework, the modification of its equation of motion outlined in Sec. \ref{SecIIB} is used. The results show that both the Lindblad and SEAQT models produce similar results for this experiment. Models to fit the constants $\gamma_1$ and $\gamma_2$ for the Lindblad equation and $\tau_{D_R}$ for the SEAQT equation of motion are used. 

\hspace{1em}
The results seen in this figure indicate that the qubit that looses information the fastest is Q1, while the loss for the other 4 qubits is significantly less. In addition, there is some deviation between the experimental results and the Lindblad and SEAQT predictions. This deviation could come from either source of coherent errors or instabilities of near-resonant two-level-systems (TLS) coupled to the qubit. This phenomenon is usually reported in these kind of devices  \cite{Burnett2019}. The Lindblad relaxation and dephasing strength parameter values and the SEAQT qubit-reservoir ($x_0\left(\tau_{D_R} \right)$) and single-qubit dephasing parameter ($\tau_{D_J}$)  values used for each qubit are shown in Table \ref{Table1} as are the characteristic experimental relaxation and dephasing times. The table also includes the detuning frequency, $\Delta f$, for each qubit. Note that the two-qubit dephasing parameter $\tau_{D_J}^{2Q}$ values are not used in this experiment but instead in the second scenario of the two-qubit entanglement gate experiment of Section \ref{SecIVC}.

\begin{table}[ht]
\begin{center}
\begin{tabular}{| c | c | c | c | c | c |}
\hline
 Parameter & Q0 & Q1 & Q2 & Q3 & Q4 \\ 
 \hline
 $x_0(\tau_{D_R}) [\mu s]$& 117.5 & 60.5 &141.3 &117.5 &130.6 \\  
 $\tau_{D_J} [\mu s]$& 40.6 & 11.3 & 43.9 & 28.1 & 49.8  \\
 $\tau_{D_J}^{2Q}  [\mu s]$ & 26.5 & 25.5 & - & - & - \\
$1/\gamma_1 [\mu s]$& 184.3 & 97.25 & 231.5 &190.2 &206.13\\
 $1/\gamma_2  [\mu s]$& 751.4 & 73.2 &637.6 &277.4 & 692.1  \\
 $T_{1} [\mu s]$ & 24.3 & 71.2& 5.9 &96.6 &100.7  \\ 
  $T_{2} [\mu s]$ & 41.9& 41.9 & 59.1& 160.5 & 171.1 \\ 
  $\Delta f [kHz]$ & 152.6 & 161.1 & 303.1 & 128.7 & 88.5 \\
  \hline
\end{tabular}
\end{center}
 \caption{\label{Table1}Summary of the simulation and experimental parameter values used for each of the $ibmq\_bogota$ qubits.}
\end{table}

\begin{figure}[htbp!]
\includegraphics[width=9cm]{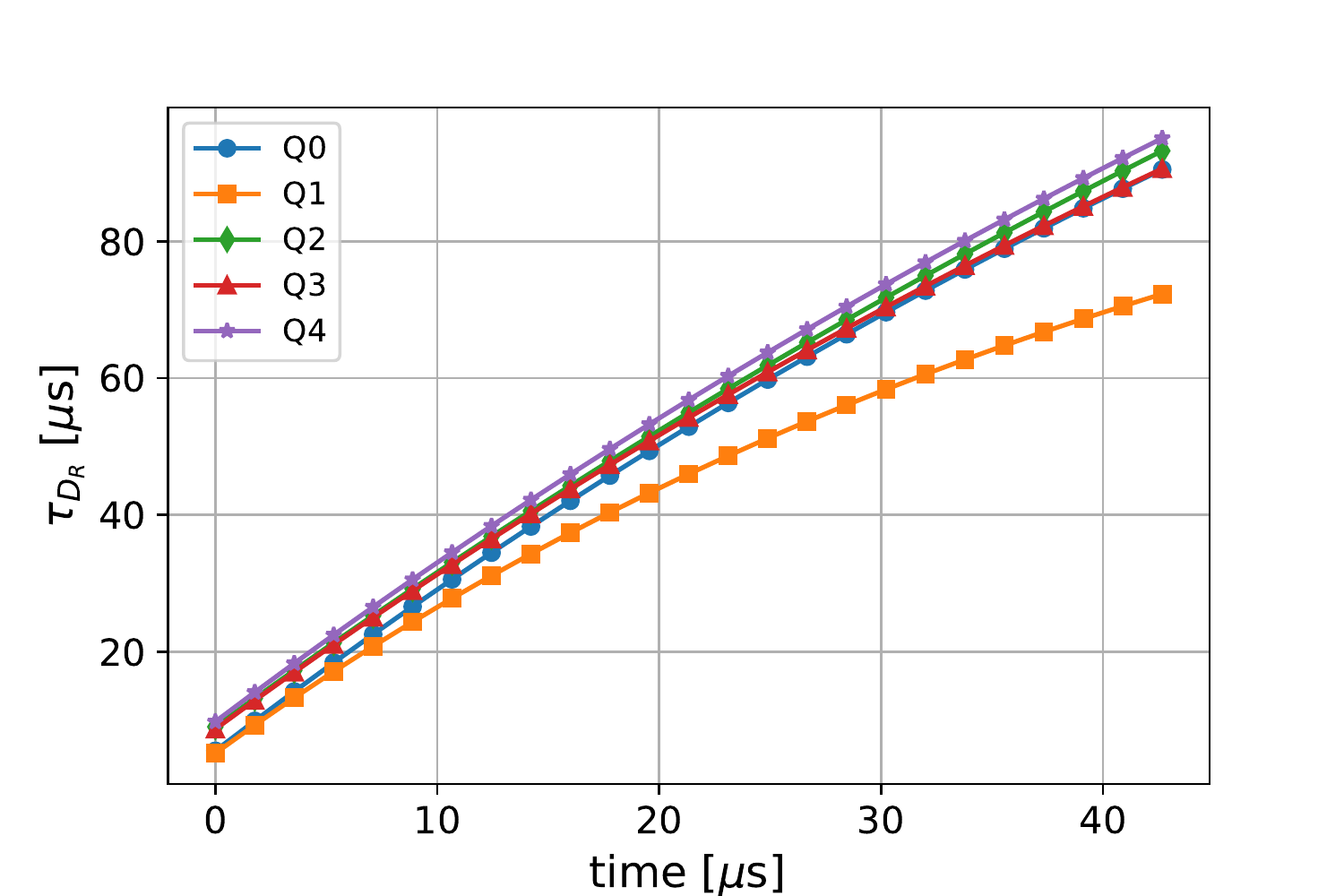}
\caption{\label{tauDRs} Evolution in time of $\tau_{D_R}$ for each qubit.}
\end{figure}

\hspace{1em}As shown in Eq. (\ref{tauDR}), the value of $\tau_{D_R}$ depends on the evolution of $\hat\rho$. As $\hat \rho$ approaches state $| 0 \rangle$, the relaxation parameter $\tau_{D_R}$ increases, which translates into a decrease in the dissipation experienced by the qubit-reservoir interaction. The evolution of $\tau_{D_R}$ for each qubit is shown in Fig. \ref{tauDRs}. As can be seen, the rate of increase of $\tau_{D_R}$ for Q1 is significantly less than that for the other qubits and as a consequence the rate of information loss in Q1 is in general greater than that for any of the other qubits as shown in Fig. \ref{t1_res}.

\subsection{Ramsey experiment}
\begin{figure}[htbp!]
\includegraphics[width=9cm]{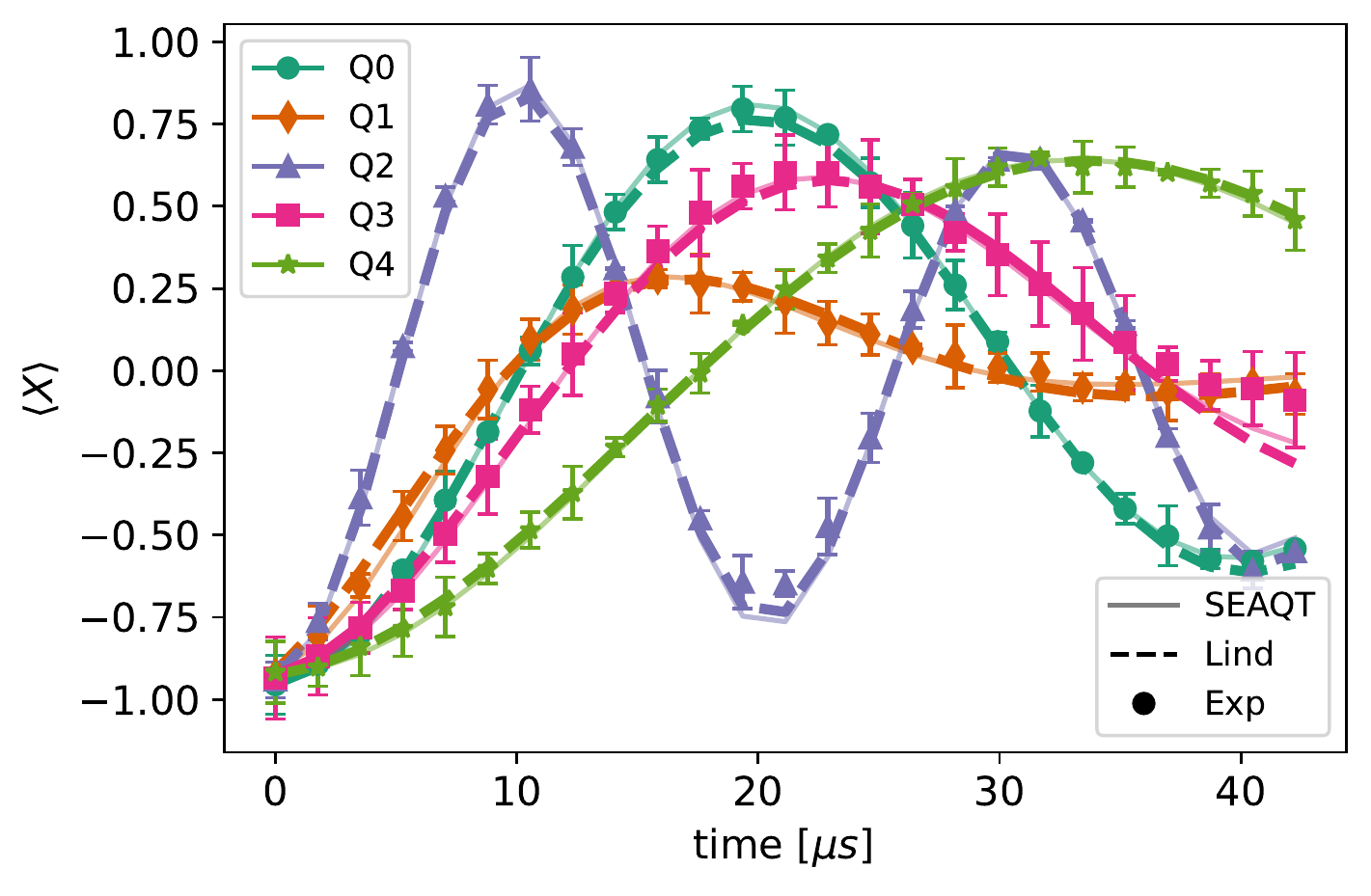}
\caption{\label{t2x} Results from the Ramsey experiment for the time evolution of the $\langle \hat X \rangle=\textrm{Tr}(\hat\rho\hat\sigma_x)$ component: the experimental results are compared with the simulation results of the SEAQT and Lindblad equations of motion for all 5 $ibmq\_bogota$ qubits. The error bars represent the standard deviation.}
\end{figure}

\hspace{1em}Fig. \ref{t2x} shows results for the Ramsey experiment of the time evolution of the $\langle \hat{X} \rangle$ component or observable. The oscillations observed in this experiment are due to the detuning frequency $\Delta f$ values given in Table \ref{Table1}. As can be seen, the amplitude of the oscillation decays with increments in the delay time for all the qubits of the {\it ibmq\_bogota}. This phenomenon called dephasing is responsible for the loss of information of the $\langle \hat X \rangle$ and $\langle \hat Y \rangle$ observables of single qubits. The average rate of decay of dephasing is quantified by the $T_2^*$ time, by $\tau_{DJ}$ for the case of the SEAQT equation of motion, and by $1/\gamma_2$ for the Lindblad equation. Values for these parameters are shown in Table \ref{Table1} for the different qubits.

\begin{figure}[htbp!]
\includegraphics[width=9cm]{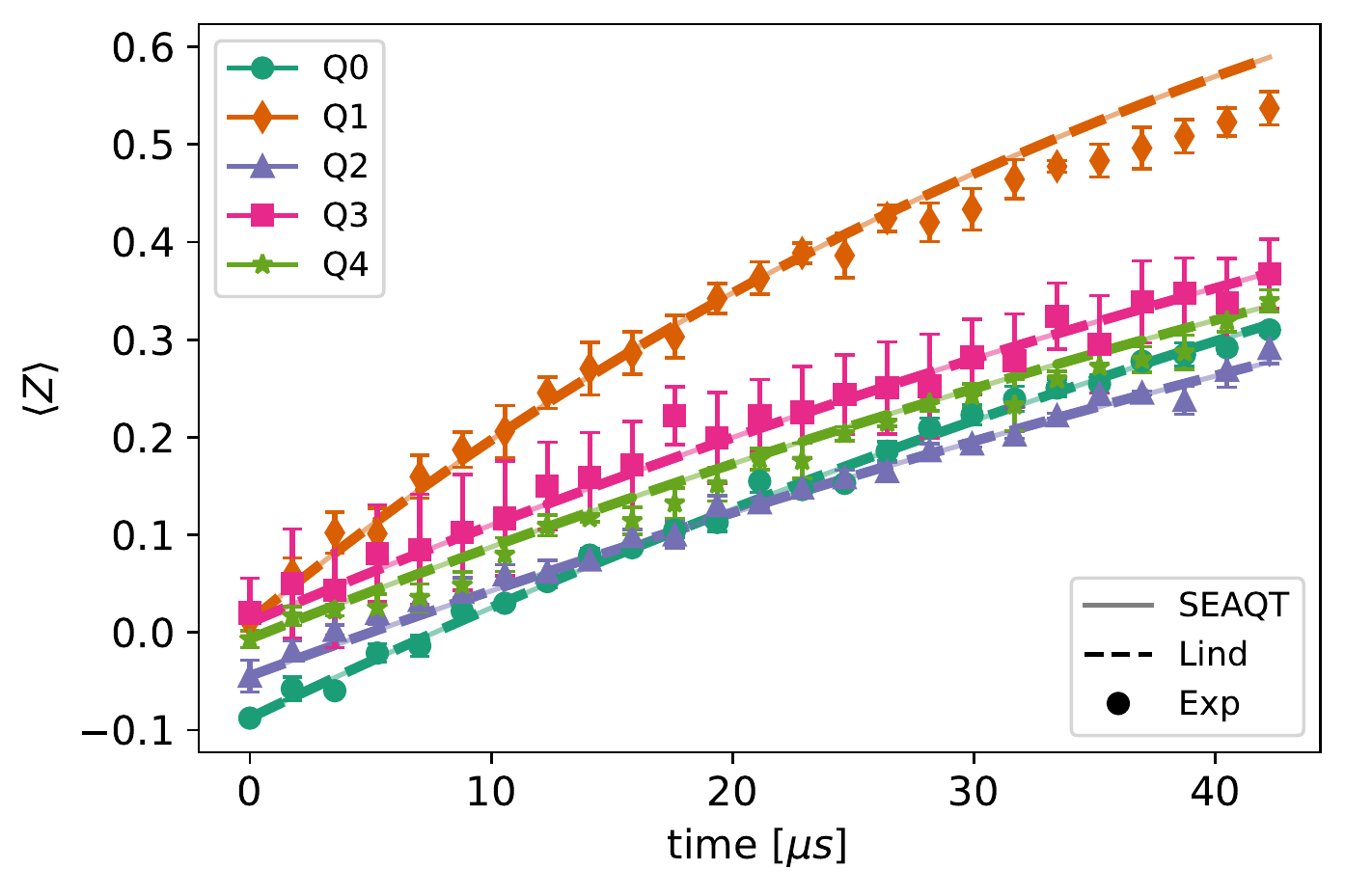}
\caption{\label{t2z} Results from the Ramsey experiment for the time evolution of the $\langle \hat Z \rangle$ component: the experimental results are compared with simulation results of the SEAQT and Lindblad equation of motion for all 5 $ibmq\_bogota$ qubits. The error bars represent the standard deviation.}
\end{figure}

\hspace{1em}In addition to the decay of the  $\langle \hat X \rangle$ observable seen in this experiment, the relaxation phenomenon resulting from an interaction with the environment (reservoir) is present. The latter's effect on the $\langle \hat{Z} \rangle$ component or observable is shown in Fig. \ref{t2z} and compared with the the SEAQT and the Lindblad predictions. In this case, the values used for $\tau_{D_R}$ for the SEAQT equation of motion and $\gamma_1$ for the Lindblad equation are those obtained for the inversion recovery experiment. As seen, predictions for both models agree quite well with the experimental results.

\subsection{Two-qubit entanglement-disentanglement experiment}\label{SecIVC}
\begin{figure}[htbp!]
a)\includegraphics[width=8cm]{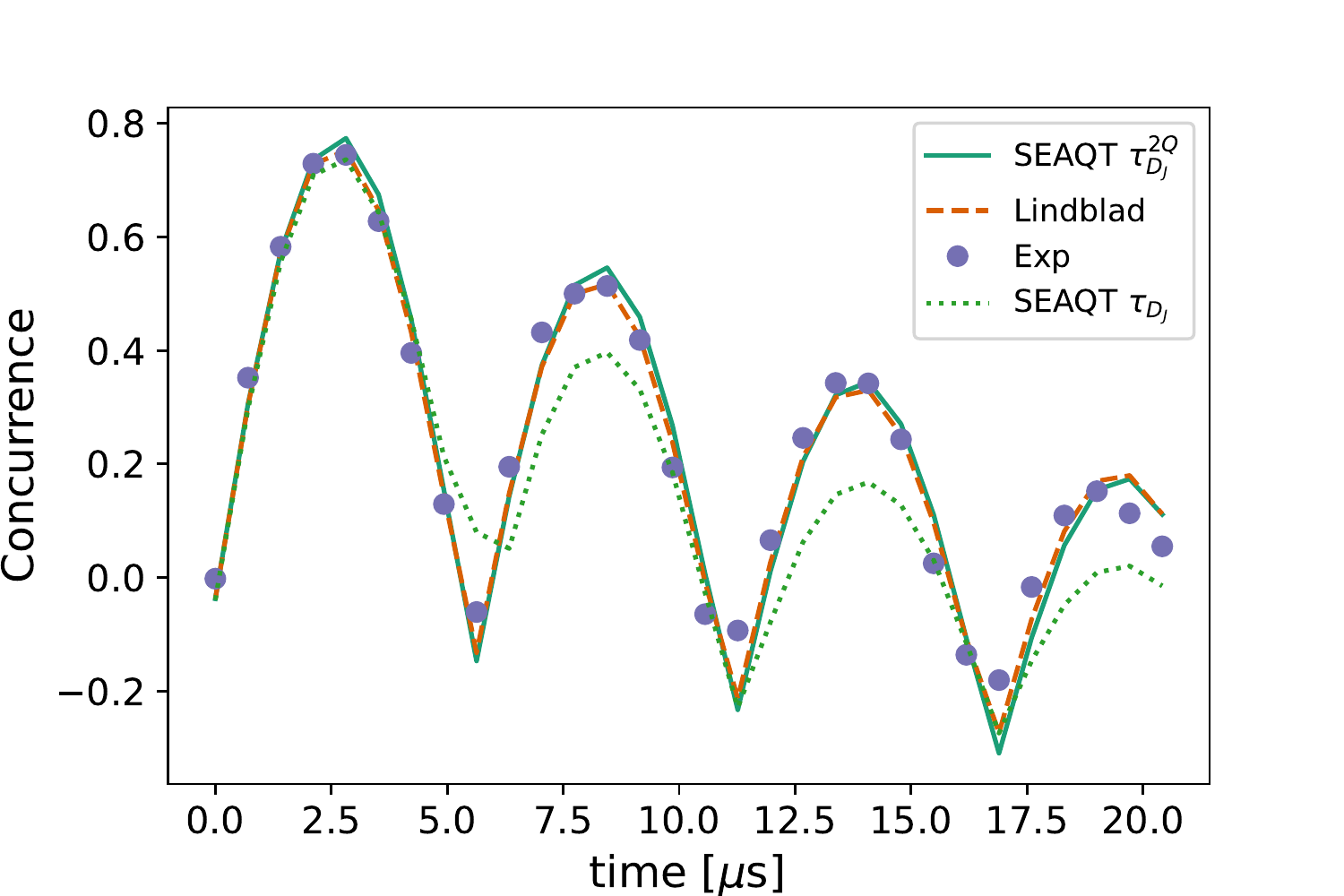}
b)\includegraphics[width=8cm]{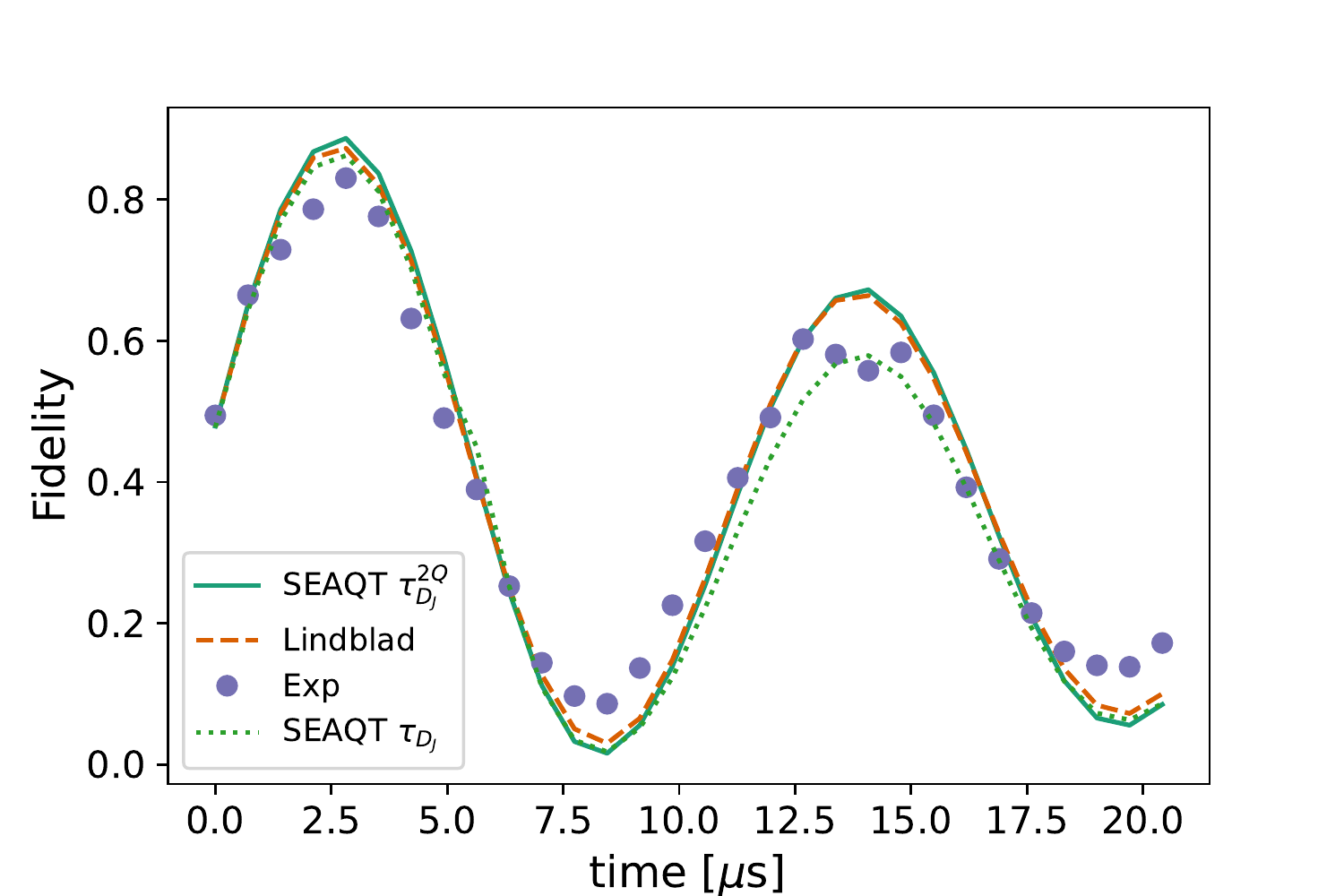}
\caption{\label{ConF} Experimental and model results for the two-qubit entanglement-disentanglement experiment showing the time evolution of a) the concurrence and b) the fidelity.}
\end{figure}

\hspace{1em}Finally, a two-qubit entanglement-disentanglement experiment between the qubits Q0 and Q1 is executed to explore the decay rate of information stored in the two-qubit system. This experiment shows how the maximum entanglement is lost with time. The results in Fig. \ref{ConF} a) show a continuous loss in the maximum concurrence as the width of the CR sequence goes from 0 to 20.45 $\mu s$. In an ideal case, the concurrence should be oscillating between 0 and 1. However, the maximum concurrence never reaches 1 and gradually decreases from a maximum value of about 0.75 to about 0.19 as the width of the CR pulse increases. The same is true for the fidelity shown in Fig. \ref{ConF} b) in which the maximum fidelity decreases from about 0.85 to about 0.68 for the second peak.

\hspace{1em}Two different scenarios are tested with the SEAQT equation of motion. The first uses the relaxation $\tau_{D_R}$ and dephasing $\tau_{DJ}$ parameter values found for the single-qubit experiments with qubits Q0 and Q1, while the second keeps the relaxation parameter $\tau_{D_R}$ values of the single-qubit experiments but utilizes values for the dephasing  parameters $\tau_{DJ}^{2Q}$ found from the present two-qubit experiment. The results for the first scenario show that the decoherence is greater {\color{red}than} what is found in the two-qubit experiment (see the dotted-green line in Fig. \ref{ConF}). In contrast, the second case, which utilizes two-qubit experimental dephasing values for $\tau_{D_J}^{2Q}$ and single-qubit experimental qubit-reservoir values for $\tau_{D_R}$, predicts the experimental concurrence and fidelity values quite well as seen in Fig. \ref{ConF} with the red-solid line. Here, the improved fit for scenario two is explained by the fact that the the $T_1$ relaxation and $T_2^*$ dephasing times characterizing the experiments change with time, a conclusion supported by Burnett et. al. \cite{Burnett2019}, who indicate that the decay times for relaxation and dephasing are not constant but vary with time. In this scenario, experiments for dephasing and relaxation were conducted on the same day, while the two-qubit experiment was taken some days later. This could influence the different decay rates. Another plausible explanation is that the disentanglement-entanglement experiment is improving the decay rate of the decoherence phenomena by the dynamics involved. Further investigation with respect to this is needed but is beyond the scope of the present work.
 
\section{Conclusions}\label{SecV}
\hspace{1em}In this paper, an approach based on the principle of steepest entropy ascent is used to predict the relaxation, dephasing, and loss of entanglement phenomena in superconductive qubits during the state evolution of inversion recovery, Ramsey, and entanglement-disentanglement experiments. The results obtained suggest that the SEAQT framework can predict the different decoherence scenarios occurring in superconductive qubits. These results supplement the purely dephasing results found previously by applying the SEAQT framework to a CPHASE gate on a double-quantum-dot qubit device \cite{Montanez-Barrera2020}. In the present paper, the phenomenon of relaxation, which requires an interaction with the environment and which was not previously addressed, is successfully modeled.  The SEAQT framework is, thus, able to effectively describe both types of phenomena. Predictions of the effects of the relaxation and dephasing phenomena have been shown to be useful in mitigating errors in NISQ devices \cite{Sun2021}. Thus, the SEAQT framework could potentially be used as the basis for an error mitigation scheme in such devices. The method to do so would be similar to the zero-noise extrapolation technique \cite{Kandala2018}, but in this case, different delay times would be used to make the extrapolation of a zero delay time such that the dephasing and relaxation noise is reduced. However, a comparison with the commonly used Lindblad equation would still be needed to determine the advantages and disadvantages of such an error mitigation technique.

\hspace{1em}Another point to make is that the use of a variable $\tau_{D_R}$ that depends on the energy of the system provides a relaxation parameter for the SEAQT equation of motion that results in predictions that compare well with the experimental data for the relaxation and Ramsey experiments. 

\hspace{1em}Clearly, the SEAQT framework is a reasonable model for predicting the dynamics of quantum protocols,
providing an alternative approach for determining $T_1$ and $T_2$ in superconductive quantum processors. As
described at the beginning of Section IIA, conceptually the SEAQT framework treats the dephasing phenomenon
as intrinsic to the system, while the open quantum system framework, which is the basis for the
Lindblad equation, treats it extrinsically, requiring two limiting assumptions: i) weak couplings with an
environment and a linear description of the non-linear evolution of the density operator. Neither of these
limitations apply to the SEAQT framework. Of course, with this loss of generality, the computational
cost of the Lindblad equation is less than that of the SEAQT equation but only slightly so. Furthermore,
the SEAQT framework’s greater generality results in an easier setup since the specific form of the environmental
interaction needed by the Lindblad equation for the dephasing phenomenon is not required
by the SEAQT equation, which treats this phenomenon intrinsically via the SEA principle. Even for the
relaxation phenomenon, a specific form of the interaction is not needed by the SEAQT equation of motion.
Of course, both the Lindblad and SEAQT equations of motion can be scaled to larger qubit arrays
and do so on the basis of $2^N$ where $N$ is the number of qubits in the array. Thus, since both equations
are 1st order ordinary differential equations in time, an array of up to at least 20 qubits could be run
on a desktop computer (e.g., an iMac). Clearly, larger arrays would require additional computational
resources although even then both approaches would in the end be limited to relatively small quantum
devices or subsets of larger devices.

\hspace{1em}Finally, the possibility of conducting experiments on cloud-based quantum devices opens new opportunities or testing different equations of motion and scenarios for the non-equilibrium evolution of quantum systems. Future work will focus on developing ways to mitigate the error inherent to these devices and recover as much of the information as is possible lost in the operation of these devices.

\begin{acknowledgments}
\vspace{-10pt}

J. A. Monta\~nez-Barrera thanks the National Council of Science and Technology (CONACyT), Mexico, for his Assistantship No. CVU-736083. S. Cano-Andrade and C.E. Damian-Ascencio gratefully acknowledge the financial support of CONACyT, Mexico, under its SNI program. The authors acknowledge the use of IBM's Quantum services for this work, the views expressed are those of the authors and do not reflect the official policy or position of IBM or the IBM Quantum team.

\end{acknowledgments}

\bibliography{References}

\section{Appendix}
\hspace{1em}For the case of superconductive qubits as in the case of the IBM $ibqm\_bogota$ device, the CNOT gate is composed of a cross resonance (CR) interaction \cite{Chow2011} and single-qubit DRAG pulses with virtual z-rotations \cite{Gambetta2011,McKay2017}. In a recent paper, Magensa \textit{et al}. \cite{Magesan2020} introduced the effective Hamiltonian of a CR interaction, i.e.,
\begin{eqnarray}\label{A24}
\hat{H}(\Omega) &=& \nu_{ZX} \frac{\hat{Z}\hat{X}}{2} + \nu_{IZ} \frac{\hat{I}\hat{Z}}{2} + \nu_{IX} \frac{\hat{I}\hat{X}}{2}  \nonumber \\
& & + \nu_{ZI}\frac{\hat{Z}\hat{I}}{2} + \nu_{ZZ} \frac{\hat{Z}\hat{Z}}{2}. \\ \nonumber
\end{eqnarray}
where $\{\hat{I}, \hat{X}, \hat{Y}, \hat{Z}\}$ are the identity and Pauli matrices. For the different tensor products (e.g., $\hat{Z}\hat{X}$), the convention is that the first acts on the control qubit and the second on the target qubit. The coefficients $\nu_{ij}$ in Eq. (\ref{A24}) are functions of the system parameters and the CR pulse amplitude $\Omega$. In this Hamiltonian, only the $\hat{Z}\hat{X}$ term, which is locally the equivalent of a CNOT gate, is of interest here. Applying the echo sequence $\hat{U} = \hat{X}\hat{I}\cdot e^{-i\hat{H}(-\Omega)t_g}\cdot \hat{X}\hat{I}\cdot e^{-i\hat{H}(\Omega)t_g}$, which as shown in Sundaresan \textit{et al.} \cite{Sundaresan2020} can be modeled by $\hat{U} = A_{II}\hat{I}\hat{I} + A_{IY}\hat{I}\hat{Y} + A_{IZ}\hat{I}\hat{Z} + A_{ZX}\hat{Z}\hat{X}$, the following Hamiltonian is obtained: 
\begin{equation}
\hat{H}_{eff} =  \tilde{\nu}_{ZX} \frac{\hat{Z}\hat{X}}{2}  +  \tilde{\nu}_{IY} \frac{\hat{I}\hat{Y}}{2} + \tilde{\nu}_{IZ} \frac{\hat{I}\hat{Z}}{2} .
\label{Heff}
\end{equation}  
Here, the $\tilde\nu_{i,j}$ are coefficients of the echo sequence effective Hamiltonian and the $\{A_{II}, A_{IY}, A_{IZ}, A_{ZX}\}$ are functions of these coefficients. If there is  crosstalk or phase misalignment, the additional rotations $\hat{Z}\hat{Y}$ and $\hat{Z}\hat{Z}$ show up in Eq. (\ref{Heff}). 

\hspace{1em}Two main strategies have been used to reduce the coherent error for the CR pulse, namely, an active cancellation on the target qubit \cite{Sheldon2016} and the addition of target rotary pulses \cite{Sundaresan2020}. These strategies reduce the error in the two-qubit subspace and even on spectator qubits, which are neighbors of the target qubit, using the approach of Sundaresan \textit{et al}.\cite{Sundaresan2020}. Both papers have a clear strategy for reducing coherent errors. First, they identify the unwanted CR Hamiltonian error terms remaining after the standard echo sequence. Second, they devise strategies to measure the error. Finally, they mitigate the error with additional pulses on the target qubit. 

\hspace{1em}Our approach is to use this calibration process for the IBM $ibqm\_bogota$ device, creating an entanglement state for different coupling factors in $\nu_{ZX}$

\end{document}